# Quantum Wells and the Generalized Uncertainty Principle

**Gardo Blado[*], Vincent Meyers and Constance Owens**

*College of Science and Mathematics*
*Houston Baptist University*
*7502 Fondren Rd., Houston, Texas, U.S.A*

## Abstract

The finite and infinite square wells are potentials typically discussed in undergraduate quantum mechanics courses. In this paper, we discuss these potentials in the light of the recent studies of the modification of the Heisenberg uncertainty principle into a generalized uncertainty principle as a consequence of attempts to formulate a quantum theory of gravity. The fundamental concepts of the minimal length scale and the generalized uncertainty principle are discussed and the modified energy eigenvalues and transmission coefficient are derived.

Key words: square well, generalized uncertainty principle, minimal length

---

[*] gblado@hbu.edu

## I. Introduction

The prediction by quantum gravity theories such as string theory and loop quantum gravity, of a minimum length scale has paved the way to discuss quantum gravity effects (which used to only be accessible to students with backgrounds in general relativity and quantum field theory) in ordinary non-relativistic quantum mechanics at the level studied in undergraduate quantum mechanics. In string theory for instance, this minimum measurable length is of the order of the dimension of the fundamental string. A number of papers have been written discussing the phenomenological consequences of the minimal length scale[1,2,3,4,5]. A considerable amount of references can be found in a recent review by S. Hossenfelder[6].

This minimal length can be shown to arise from the modification of the Heisenberg uncertainty principle (UP) to a generalized uncertainty principle (GUP). The UP is given by $\Delta x_i \Delta p_i \geq \frac{\hbar}{2}$ [†] where the index $i = 1, 2, 3$ represent the $x$, $y$, and $z$ directions of the position and momentum of a particle. This UP can be derived from the more general uncertainty inequality for any two observables $A$ and $B$ given by[7] $(\Delta A)^2 (\Delta B)^2 \geq \left(\frac{1}{2i}\langle[\hat{A},\hat{B}]\rangle\right)$ by using the commutation relation of the position and momentum operators, $[\hat{x}_i, \hat{p}_i] = i\hbar \delta_{ij}$. Quantum gravity theories however, modify this commutation relation to[3]

$$[\hat{x}_i, \hat{p}_j] = i\hbar[\delta_{ij} + \beta\delta_{ij}p^2 + 2\beta p_i p_j] \quad (1)$$

where $\beta$ is a small parameter which depends on the Planck length and Planck's constant. This modification in the commutator gives rise to a GUP[3]

$$\Delta x_i \Delta p_i \geq \frac{\hbar}{2}[1 + \beta((\Delta p)^2 + \langle p \rangle^2) + 2\beta(\Delta p_i^2 + \langle p_i \rangle^2)] \quad (2)$$

Confining the discussion in one dimension, we see that for the UP, $\Delta x \geq \frac{\hbar}{2\Delta p}$. Hence, one can in principle decrease $\Delta x$ arbitrarily by increasing $\Delta p$. We show this behavior in Figure 1 which gives a schematic plot of $\Delta x \sim 1/\Delta p$.

---

[†] A "tighter" form of the uncertainty principle is given by $(\Delta q)^2 (\Delta p)^2 - \langle\frac{\{(q-\langle q\rangle),(p-\langle p\rangle)\}}{2}\rangle^2 \geq \frac{\hbar^2}{4}$ with $\langle q \rangle = \langle\psi|q|\psi\rangle$. See reference [14].

However for the GUP in equation (2), $\Delta x \sim \frac{\hbar}{2\Delta p} + \frac{\hbar\beta}{2}\Delta p$, and increasing $\Delta p$ does not arbitrarily decrease $\Delta x$ because the term $\frac{\hbar\beta}{2}\Delta p$ dominates. Instead, this yields a minimum value for $\Delta x = \Delta x_0$, which is the minimum length. This is shown schematically in Figure 2 by the graph of $\Delta x \sim 1/\Delta p + \Delta p$. It can be shown that equation (1) can be realized by setting[3]

$$x_i = x_{0i}, p_i = p_{0i}(1 + \beta p_0^2) \tag{3}$$

with $p_0^2 = \sum_{j=1}^{3} p_{0j} p_{0j}$ with $x_{0i}$ and $p_{0i}$ satisfying the usual commutation relation $[\hat{x}_{0i}, \hat{p}_{0j}] = i\hbar\delta_{ij}$ and with $\hat{p}_{0i} = \frac{\hbar}{i}\frac{d}{dx_i}$.

In this paper we will illustrate how equation (3) will modify the quantum mechanics of the one-dimensional finite square well and the infinite square well. We chose to discuss these familiar quantum wells because they are discussed in a typical undergraduate quantum mechanics class.

In section II, we discuss how the GUP modifies the Hamiltonian in the time independent Schrodinger equation. In section III, we review the known results for the finite square well (FSW) using the approach of Goswami[8] but using the complex exponential solution all throughout. We use the approach of section III to derive the GUP-modified results in section IV since it is easier to solve the fourth order differential equation of equation (7) using exponentials when the GUP effects are taken into account. In section V, we briefly discuss the GUP effect on the infinite square well potential. Some conclusions are discussed in section VI.

## II. Gravitational Effects in Quantum Mechanics

Given a particle of mass $m$ subjected to a potential $V(x)$, one can describe its quantum mechanical behavior by solving the wavefunctions $\psi(x)$ and its energies $E$ from the time independent Schrodinger equation (TISE)

$$\hat{H}\psi = E\psi \Rightarrow \left(\frac{\hat{p}^2}{2m} + V\right)\psi = E\psi \tag{4}$$

where $\hat{H}$ is the Hamiltonian operator given by $\frac{\hat{p}^2}{2m} + V$. In ordinary quantum mechanics,

$$\hat{p} = \frac{\hbar}{i}\frac{d}{dx}. \qquad (5)$$

Plugging in equation (5) into equation (4),

$$E\psi = \left(\frac{1}{2m}\left(\frac{\hbar}{i}\frac{d}{dx}\right)^2 + V\right)\psi = -\frac{\hbar^2}{2m}\frac{d^2\psi}{dx^2} + V\psi \qquad (6)$$

which is the familiar form of the TISE in ordinary quantum mechanics. The Schrodinger equation is modified by the GUP by substituting equation (3) with $\hat{p}_0 = \frac{\hbar}{i}\frac{d}{dx}$ into equation (4). Keeping only terms up to the order of $\beta$, we get

$$E\psi = \left(\frac{p_0^2}{2m}(1+\beta p_0^2)^2 + V\right)\psi = \left(\frac{p_0^2}{2m}(1+2\beta p_0^2) + V\right)\psi \quad \text{or} \quad \left(\frac{p_0^2}{2m} + \frac{\beta}{m}p_0^4 + V\right)\psi = E\psi.$$

Substituting $\hat{p}_0 = \frac{\hbar}{i}\frac{d}{dx}$ into the preceding equation gives

$$E\psi = \left(-\frac{\hbar^2}{2m}\frac{d^2}{dx^2} + \frac{\beta\hbar^4}{m}\frac{d^4}{dx^4} + V\right)\psi. \qquad (7)$$

Equation (7) is the GUP version of equation (6) with the extra term $\frac{\beta\hbar^4}{m}\frac{d^4\psi}{dx^4}$. It is interesting to note that the extra term is similar to the first correction (quartic in $p$) that arises when the classical kinetic energy term is replaced by the special relativistic kinetic energy expression in the time independent Schrodinger equation (see page 268 of reference [7]).

## III. Finite Square Well

We review the solution of the Finite Square Well. The approach in this section will be used in deriving our results with GUP.

The FSW potential is given by the potential

$$V(x) = \begin{cases} -V_0 & \text{for } -a < x < a \\ 0 & \text{for } |x| > a \end{cases} \qquad (8)$$

Solving the Schrodinger equation $-\frac{\hbar^2}{2m}\frac{d^2\psi}{dx^2} + V(x) = \mathcal{E}\psi$ for the <u>bound states</u> with $\mathcal{E} = -E$ where $V_0 > E > 0$, we get the following physical solutions

$$\begin{cases} \psi_1(x) = Be^{kx} \\ \psi_2(x) = De^{ilx} + Je^{-ilx} \\ \psi_3(x) = Ge^{-kx} \end{cases}$$

where $\psi_1(x)$, $\psi_2(x)$, and $\psi_3(x)$ are the solutions for the regions $x < -a, -a < x < a$, and $x > a$ respectively, $k = \sqrt{\frac{2mE}{\hbar^2}}$ and $l = \sqrt{\frac{2m}{\hbar^2}(V_0 - E)}$. Applying the boundary conditions,

$$\begin{cases} \psi_1(-a) = \psi_2(-a) \\ \psi_2(a) = \psi_3(a) \\ \left.\frac{d\psi_1}{dx}\right|_{x=-a} = \left.\frac{d\psi_2}{dx}\right|_{x=-a} \\ \left.\frac{d\psi_2}{dx}\right|_{x=a} = \left.\frac{d\psi_3}{dx}\right|_{x=a} \end{cases} \quad (9)$$

we get the following system of equations.

$$\begin{cases} Be^{-ka} - De^{-ila} - Je^{ila} = 0 \\ Bke^{-ka} - iDle^{-ila} + iJle^{ila} = 0 \\ De^{ila} + Je^{-ila} - Ge^{-ka} = 0 \\ iDle^{ila} - iJle^{-ila} + Gke^{-ka} = 0 \end{cases}$$

Solving this system of equations for the constants $B$, $D$, $J$, and $G$, yields a trivial solution in which the constants are zero which yields non-normalizable solutions. Hence we set the determinant of the matrix with elements consisting of the coefficients of $B$, $D$, $J$, and $G$, to zero. This yields $e^{4ila} = \frac{l^2 - k^2 + 2ilk}{l^2 - k^2 - 2ilk}$. Using the formula, $e^{i\theta} = \cos\theta + i\sin\theta$, we get from the preceding equation $\tan 2la = \frac{2lk}{l^2 - k^2}$. Using the trigonometric identity $\tan 2\gamma = \frac{2\tan\gamma}{1-\tan^2\gamma}$ we get

$\boldsymbol{tanla = k/l}$ and $-l/k$ or $\boldsymbol{tanla = k/l}$ (even solutions) and $\boldsymbol{cotla = -k/l}$ (odd (10) solutions)

which are the well-known results found in textbooks[7].

Turning our attention to the scattering states with $\mathcal{E} = E$ where $E > 0$, we get the following physical solutions, assuming a particle incident from the left ($x < -a$)

$$\begin{cases} \psi_1(x) = Ae^{ikx} + Be^{-ikx} \\ \psi_2(x) = Ce^{i\sigma x} + De^{-i\sigma x} \\ \psi_3(x) = Fe^{ikx} \end{cases}$$

where $\psi_1(x)$, $\psi_2(x)$, and $\psi_3(x)$ are the solutions for the regions $x < -a$, $-a < x < a$, and $x > a$ respectively, $k = \sqrt{\frac{2mE}{\hbar^2}}$ and $\sigma = \sqrt{\frac{2m}{\hbar^2}(V_0 + E)}$. Applying once more the boundary conditions in equation (9), we get the following system of equations.

$$\begin{cases} Ae^{-ika} + Be^{ika} - Ce^{-i\sigma a} - De^{i\sigma a} = 0 \\ iAke^{-ika} - iBke^{ika} - iC\sigma e^{-i\sigma a} + iD\sigma e^{i\sigma a} = 0 \\ Ce^{i\sigma a} + De^{-i\sigma a} - Fe^{ika} = 0 \\ iC\sigma e^{i\sigma a} - iD\sigma e^{-i\sigma a} - iFke^{ika} = 0 \end{cases}$$

One can solve for the constants $B$, $C$, $D$, and $F$ in terms of $A$ in the above equation. Computing for the transmission coefficient $T = \left(\frac{F}{A}\right)^* \left(\frac{F}{A}\right)$, we get the following expression.

$$T = -\sigma^2 k^2 \{2k^2\sigma^2 \cos^2(\sigma a) - 2k^2\sigma^2\cos^4(\sigma a) - k^2\sigma^2 + k^4\cos^4(\sigma a) \qquad (11)$$
$$+ \sigma^4\cos^4(\sigma a) - k^4\cos^2(\sigma a) - \sigma^4\cos^2(\sigma a)\}^{-1}$$

Plugging in $k$ and $\sigma$ in the above equation, and using some trigonometric identities, we get

$$T = \left\{1 + \frac{V_0^2}{4E(V_0 + E)} \cdot \sin^2\left(\frac{2\sqrt{2m(V_0+E)}}{\hbar}a\right)\right\}^{-1} \qquad (12)$$

which is the well-known result found in textbooks[7].

### IV.    Finite Square Well with the Generalized Uncertainty Principle (GUP)

With the modified Schrodinger equation due to GUP, we now have to solve equation (7) with $V(x)$ given by equation (8). For the <u>bound states</u> with $\mathcal{E} = -E$ where $V_0 > E > 0$, we get the following physical solutions,

$$\begin{cases} \psi_1(x) = Be^{k'x} \\ \psi_2(x) = Fe^{il'x} + Ge^{-il'x} \\ \psi_3(x) = Je^{-k'x} \end{cases} \qquad (13)$$

where $\psi_1(x)$, $\psi_2(x)$, and $\psi_3(x)$ are the solutions for the regions $x < -a, -a < x < a$, and $x > a$ respectively, $k' = k + \hbar^2 k^3 \beta$ and $l' = l - \hbar^2 l^3 \beta$. The appendix explains in detail the derivation of equation (13) above. Applying the boundary conditions of equation (9), we get the following system of equations.

$$\begin{cases} Be^{-k'a} - Fe^{-il'a} - Ge^{il'a} = 0 \\ Bk'e^{-k'a} - iFl'e^{-il'a} + iGl'e^{il'a} = 0 \\ Fe^{il'a} + Ge^{-il'a} - Je^{-k'a} = 0 \\ iFl'e^{il'a} - iGl'e^{-il'a} + Jk'e^{-k'a} = 0 \end{cases}$$

As before, solving this system of equations for the constants B, F, G, and J, yields a trivial solution in which the constants are zero which yields non-normalizable solutions. Hence we set the determinant of the matrix with elements consisting of the coefficients of B, F, G, and J, to zero. This yields $e^{-4ail(-1+\hbar^2 l^2 \beta)} = \frac{(il-k)(2i\hbar^2 l^3 \beta - il + k + 2\hbar^2 k^3 \beta)}{(il+k)(2i\hbar^2 l^3 \beta - il - k - 2\hbar^2 k^3 \beta)}$. Using the formula, $e^{i\theta} = \cos\theta + i\sin\theta$, we get from the preceding equation, $\tan(2al(-1 + \hbar^2 l^2 \beta)) = \frac{2lk(-\hbar^2 l^2 \beta + 1 + \hbar^2 k^2 \beta)}{k^2 + \hbar^2 k^4 \beta + 2l^4 \hbar^2 \beta - l^2}$.

Using the trigonometric identity $\tan 2\gamma = \frac{2\tan\gamma}{1-\tan^2\gamma}$ we get

$$\tan(al(1 - \hbar^2 l^2 \beta)) = -\frac{l}{k}(1 - \hbar^2 \beta(k^2 + l^2)) \text{ (odd solutions) and } \tan(al(1 - \hbar^2 l^2 \beta)) = \frac{k}{l}(1 + \hbar^2 \beta(k^2 + l^2)) \text{ (even solutions)}. \quad (14)$$

It is important to note from the previous equations that the equations in (14) reduce to the non-GUP case in equation (10), when β goes to zero. The preceding results agree with the paper by Vahedi[5] which used a different approach less familiar to students in undergraduate quantum mechanics class.

To see the relative behavior of the energy eigenvalues, we plot both sides of equation (14) in Figure 3 and Figure 4 (with the energy in the horizontal axis) using Planck units with $\hbar = 1, V_0 = 8, a = \pi$ and $m = 1/2$. The energies are given by the intersections of the graphs (shown by the red and blue dots). The $\beta = 0$ graphs (blue graphs) correspond to the usual finite square well solutions (non-GUP) while the $\beta = 0.01$ (red graph) corresponds to the GUP result. It is clear that the energy eigenvalues are shifted in the GUP solutions.

Looking next at the underlined scattering states, with $\mathcal{E} = E$ where $E > 0$, we get the following physical solutions, assuming a particle incident from the left ($x < -a$)

$$\begin{cases} \psi_1(x) = Ae^{ik''x} + Be^{-ik''x} \\ \psi_2(x) = He^{i\sigma'x} + Je^{-i\sigma'x} \\ \quad\quad \psi_3(x) = Ce^{ik''x} \end{cases} \quad (15)$$

where $\psi_1(x)$, $\psi_2(x)$, and $\psi_3(x)$ are the solutions for the regions $x < -a$, $-a < x < a$, and $x > a$ respectively, $k'' = k - k^3\hbar^2\beta$ and $\sigma' = \sigma - \sigma^3\hbar^2\beta$. The appendix explains in detail the derivation of equation (15) above. Applying once more the boundary conditions of equation (9), we get the following system of equations

$$\begin{cases} Ae^{-ik''a} + Be^{ik''a} = He^{-i\sigma'a} + Je^{i\sigma'a} \\ iAk''e^{-ik''a} - iBk''e^{ik''a} = iH\sigma'e^{-i\sigma'a} - iJ\sigma'e^{i\sigma'a} \\ He^{i\sigma'a} + Je^{-i\sigma'a} = Ce^{ik''a} \\ iH\sigma'e^{i\sigma'a} - iJ\sigma'e^{-i\sigma'a} = iCk''e^{ik''a} \end{cases}$$

One can solve for the coefficients $B$, $H$, $J$, and $C$ in terms of $A$ in the above equation. Computing for the transmission coefficient $T_{GUP} = \left(\frac{C}{A}\right)^*\left(\frac{C}{A}\right)$, we get the following equation.

$$T_{GUP} = -(\sigma')^2(k'')^2\{-2(k'')^2(\sigma')^2\cos^4(\sigma'a) - (k'')^2(\sigma')^2 \quad (16)$$
$$+ 2(k'')^2(\sigma')^2\cos^2(\sigma'a) - \cos^2(\sigma'a)(k'')^4 + \cos^4(\sigma'a)(k'')^4$$
$$- \cos^2(\sigma'a)(\sigma')^4 + \cos^4(\sigma'a)(\sigma')^4\}^{-1}$$

It is clear from the above equation that we get the non-GUP result in equation (11) since $\sigma' \to \sigma$ and $k'' \to k$ for $\beta = 0$. We compare $T$ and $T_{GUP}$ in Figure 5, where we plot the transmission coefficients using equation (16) (with the energy in the horizontal axis) using Planck units with $\hbar = 1, V_0 = 8, a = \pi$ and $m = 1/2$. The $\beta = 0$ graphs (blue graph) correspond to the usual finite square well transmission coefficient[7] (non-GUP) while the $\beta = 0.01$ (red graph) corresponds to the GUP result. The energies at which the transmission coefficient are equal to one are shifted in the GUP result.

## V. Infinite Square Well with the Generalized Uncertainty Principle

As another example of the effect of the generalized uncertainty principle, let us discuss briefly the infinite square well or the particle-in-a-box potential. The potential is given by

$$V(x) = \begin{cases} 0 & \text{for } 0 < x < a \\ \infty & \text{otherwise} \end{cases}.$$

Of course, $\psi = 0$ outside $0 < x < a$ since $V$ is infinite in these regions. For the region <u>inside the box</u>, $0 < x < a$, in which $V = 0$, we have to solve (from equation (7)) $-\frac{\hbar^2}{2m}\frac{d^2\psi}{dx^2} + \beta\frac{\hbar^4}{m}\frac{d^4\psi}{dx^4} = E\psi$. The physical solutions are given by (we employ the same method as in Regions I and III of the finite square well scattering states which is explained in the appendix),

$$\psi(x) = Ae^{ik''x} + Be^{-ik''x} \tag{17}$$

with

$$k'' = k - \hbar^2 k^3 \beta \text{ and } k = \sqrt{\frac{2mE}{\hbar^2}}. \tag{18}$$

From the boundary condition $\psi(0) = 0$, we get $A = -B$. Equation (17) becomes

$$\psi(x) = Ae^{ik''x} - Ae^{-ik''x} = A(e^{ik''x} - e^{-ik''x}) = 2iA\sin k''x \tag{19}$$

while the boundary condition $\psi(a) = 0$ applied to equation (19), gives $k'' = k - \hbar^2 k^3 \beta = \frac{n\pi}{a}$ where $n = 1, 2, 3,...$ Using the value of $k$ in equation (18), we get the energies $E$ to be (up to $\mathcal{O}(\beta)$)

$$E_n = \frac{n^2\pi^2\hbar^2}{2ma^2} + \beta\frac{n^4\pi^4\hbar^4}{ma^4} \tag{20}$$

which agrees with a previous result[4]. Equation (20) clearly reduces to the ordinary (non-GUP) energies of the particle-in-a-box when $\beta = 0$.

## VI. Conclusions:

This paper demonstrates that quantum gravity effects can be discussed at the level of ordinary quantum mechanics without the need for a background in general relativity and quantum field theory. By a straightforward revision of the time-independent Schrodinger equation using the modified momentum operator in equation (3), one can solve for the shifted energies of the bound states of the finite and infinite square wells and the shifted values of the transmission coefficient of the scattering states of the finite square well.

It is not surprising that several papers have been written discussing GUP effects in potentials typically discussed in undergraduate quantum mechanics classes such as the infinite square well[1,4], free particle[9], harmonic oscillator[10,11], and the hydrogen atom[12]. Reference [3] also discussed the potential step and potential barrier functions. Besides describing the quantum mechanics of point-like objects, the minimal length can also be applied in describing non-pointlike systems such as nucleons, quasi-particles and collective excitations. Reference [15] works out the application of the minimal length to a non-pointlike particle in a d-dimensional isotropic harmonic oscillator. Another interesting potential that will be the subject of a future paper is the double square well potential[13], where tunneling is an important phenomenon. The methods discussed here can be readily applied to this potential. We believe that research on the phenomenological consequences of the minimal length is accessible to undergraduate students with a background in quantum mechanics.

# Appendix:

In this appendix, we give the details of the derivation of (13) and (15). We will call regions I, II, and III, $x < -a$, $-a < x < a$, and $x > a$ respectively.

We start with the bound states. For regions I & III, $V(x) = 0$ and since we are looking at bound states, we let $E \to -E$ taking $E$ here (i.e. in $-E$) as positive. Equation (7) becomes

$$-E\psi = -\frac{\hbar^2}{2m}\frac{d^2\psi}{dx^2} + \frac{\beta\hbar^4}{m}\frac{d^4\psi}{dx^4}.$$

The above equation can be rearranged as

$$\frac{d^2\psi}{dx^2} - j^2\frac{d^4\psi}{dx^4} - k^2\psi = 0 \tag{21}$$

where

$$j = \sqrt{2\hbar^2\beta}, \quad k = \sqrt{\frac{2mE}{\hbar^2}}$$

Following reference [3], we assume a solution of the form $\psi = e^{nx}$ where $n$ is a constant. Plugging in this solution to equation (21), we get a quartic equation in $n$ given by

$$n^2 - j^2 n^4 - k^2 = 0$$

whose solutions are $n = \left(\pm\frac{1}{\hbar\sqrt{2\beta}}, \pm(k + \beta\hbar^2 k^3)\right)$ where we only expanded up to the order of $\beta$. Hence the solution for Regions I & III is (ignoring first the coefficients) $\psi \sim e^{\frac{1}{\hbar\sqrt{2\beta}}x} + e^{-\frac{1}{\hbar\sqrt{2\beta}}x} + e^{(k+\beta\hbar^2 k^3)x} + e^{-(k+\beta\hbar^2 k^3)x}$. We drop the first 2 terms since $\psi$ does not reduce to the FSW solution when $\beta \to 0$ (non-GUP case) for both regions. For region I, we drop the term $e^{-(k+\beta\hbar^2 k^3)x}$ since this goes to infinity as $x \to -\infty$. Hence for region I, the solution is $\psi_1(x) = Be^{k'x}$ where $k' = k + \hbar^2 k^3 \beta$. For region III, we drop the term $e^{(k+\beta\hbar^2 k^3)x}$ since this goes to infinity as $x \to +\infty$. Hence for region III, the solution is $\psi_3(x) = Je^{-k'x}$.

For region II, $V(x) = -V_0$ and as before, we let $E \to -E$ with $E$ here as positive for the bound states. Equation (7) becomes

$$-E\psi = -\frac{\hbar^2}{2m}\frac{d^2\psi}{dx^2} + \frac{\beta\hbar^4}{m}\frac{d^4\psi}{dx^4} - V_0.$$

This equation can be rearranged as

$$\frac{d^2\psi}{dx^2} - j^2\frac{d^4\psi}{dx^4} + l^2\psi = 0 \tag{22}$$

with

$$l = \sqrt{\frac{2m(V_0-E)}{\hbar^2}}.$$

As before, we plug in the assumed solution $\psi = e^{nx}$ where $n$ is a constant, into equation (22). This gives us the quartic equation in $n$

$$n^2 - j^2 n^4 + l^2 = 0$$

whose solutions are $n = \left(\pm\frac{1}{\hbar\sqrt{2\beta}}, \pm i(l - \beta\hbar^2 l^3)\right)$ where we only expanded up to the order of $\beta$. We drop the first 2 solutions since $\psi$ does not reduce to the FSW solution when $\beta \to 0$ (non-GUP case). Hence for region II, the solution is $\psi_2(x) = Fe^{il'x} + Ge^{-il'x}$.

For the scattering states with $E > 0$ in equation (7), we have $E\psi = \left(-\frac{\hbar^2}{2m}\frac{d^2}{dx^2} + \frac{\beta\hbar^4}{m}\frac{d^4}{dx^4} + V\right)\psi$

$$-\frac{\hbar^2}{2m}\frac{d^2\psi}{dx^2} + \frac{\beta\hbar^4}{m}\frac{d^4\psi}{dx^4} + V\psi = E\psi \tag{23}$$

For regions I & III, $V(x) = 0$ and we get from (23),

$$-\frac{\hbar^2}{2m}\frac{d^2\psi}{dx^2} + \frac{\beta\hbar^4}{m}\frac{d^4\psi}{dx^4} = E\psi$$

Substituting $\psi = e^{nx}$ where $n$ is a constant into the preceding equation, we get a quartic equation for $n$.

$$-\hbar^2 n^2 + 2\beta\hbar^4 n^4 - 2mE = 0$$

whose solutions are $n = \left(\pm \frac{1}{\hbar\sqrt{2\beta}}, \pm i(k - \beta\hbar^2 k^3)\right)$ where we only expanded up to the order of $\beta$. Let us assume a particle incident from the left moving towards the right. Hence the solution for Regions I & III is (ignoring first the coefficients) $\psi \sim e^{\frac{1}{\hbar\sqrt{2\beta}}x} + e^{-\frac{1}{\hbar\sqrt{2\beta}}x} + e^{i(k-\beta\hbar^2 k^3)x} + e^{-i(k-\beta\hbar^2 k^3)x}$. We drop the first 2 terms since $\psi$ does not reduce to the FSW solution when $\beta \to 0$ (non-GUP case) for both regions. For region I, (with $k'' = k - k^3\hbar^2\beta$) the solution is $\psi_1(x) = Ae^{ik''x} + Be^{-ik''x}$ while for region III, the solution is $\psi_3(x) = Ce^{ik''x}$ for a right moving wave.

For region II, $V(x) = -V_0$ and equation (23) gives

$$-\frac{\hbar^2}{2m}\frac{d^2\psi}{dx^2} + \frac{\beta\hbar^4}{m}\frac{d^4\psi}{dx^4} - V_0\psi = E\psi$$

Substituting $\psi = e^{nx}$ where $n$ is a constant into the preceding equation, we get a quartic equation for $n$.

$$-\hbar^2 n^2 + 2\beta\hbar^4 n^4 - 2m(V_0 + E) = 0$$

whose solutions are $n = \left(\pm \frac{1}{\hbar\sqrt{2\beta}}, \pm i(\sigma - \beta\hbar^2\sigma^3)\right)$ where we only expanded up to the order of $\beta$ with $\sigma = \sqrt{\frac{2m}{\hbar^2}(V_0 + E)}$. The solution will be of the form $\psi \sim e^{\frac{1}{\hbar\sqrt{2\beta}}x} + e^{-\frac{1}{\hbar\sqrt{2\beta}}x} + e^{i(\sigma-\beta\hbar^2\sigma^3)x} + e^{-i(\sigma-\beta\hbar^2\sigma^3)x}$. We drop the first 2 terms since $\psi$ does not reduce to the FSW solution when $\beta \to 0$ (non-GUP case). Hence for region II, the solution is $\psi_2(x) = He^{i\sigma'x} + Je^{-i\sigma'x}$ where $\sigma' = \sigma - \beta\hbar^2\sigma^3$.

# Figures:

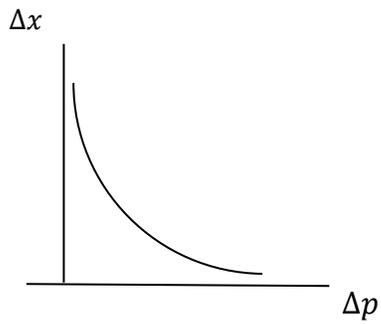

**Figure 1:** A plot of $\Delta x \sim 1/\Delta p$ in which $\Delta x$ decreases arbitrarily with increasing $\Delta p$.

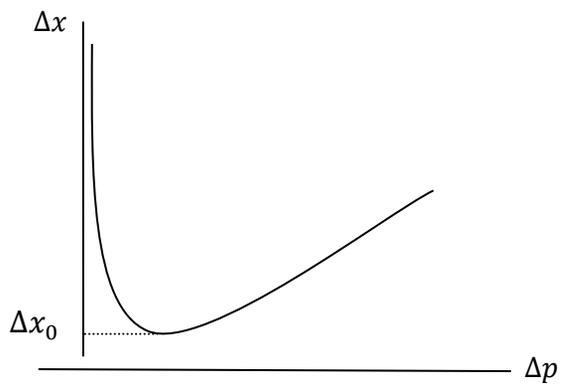

**Figure 2:** A plot of $\Delta x \sim 1/\Delta p + \Delta p$ which yields a minimum length $\Delta x_0$.

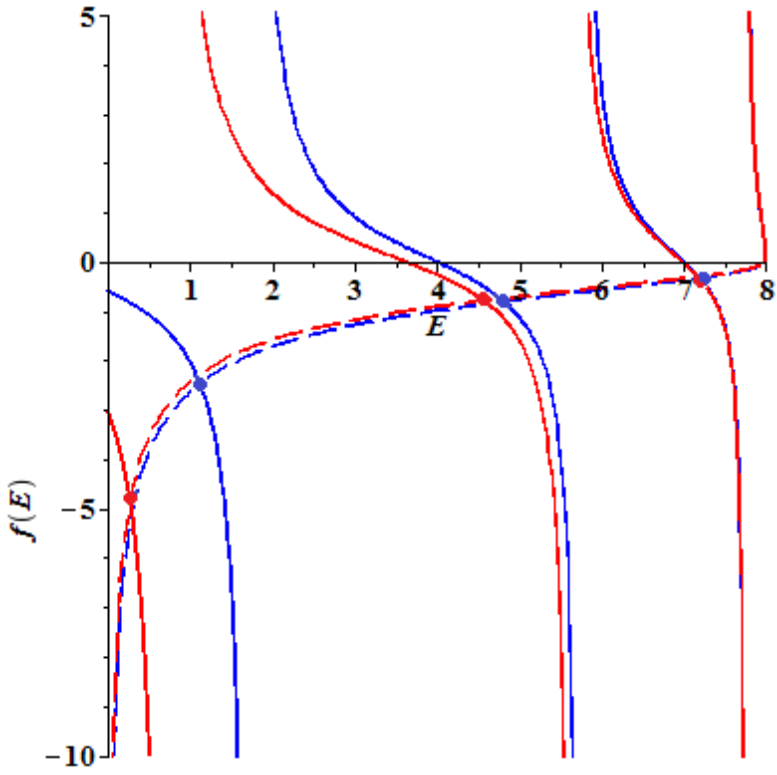

**Figure 3:** Graphs of equation (14) for $f(E) = \tan(al(1 - \hbar^2 l^2 \beta))$ and $-\frac{l}{k}(1 - \hbar^2 \beta(k^2 + l^2))$ with $\beta = 0$ (blue) and $\beta = 0.01$ (red), [odd solutions]

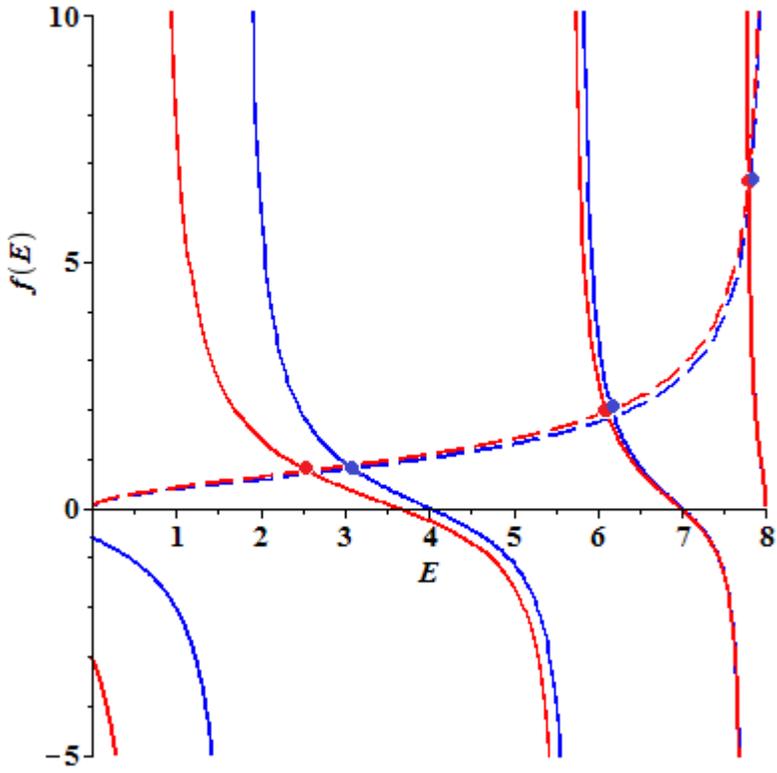

**Figure 4:** Graphs of equation (14) for $f(E) = \tan(al(1 - \hbar^2 l^2 \beta))$ and $\frac{k}{l}(1 + \hbar^2 \beta(k^2 + l^2))$ with $\beta = 0$ (blue) and $\beta = 0.01$ (red), [even solutions]

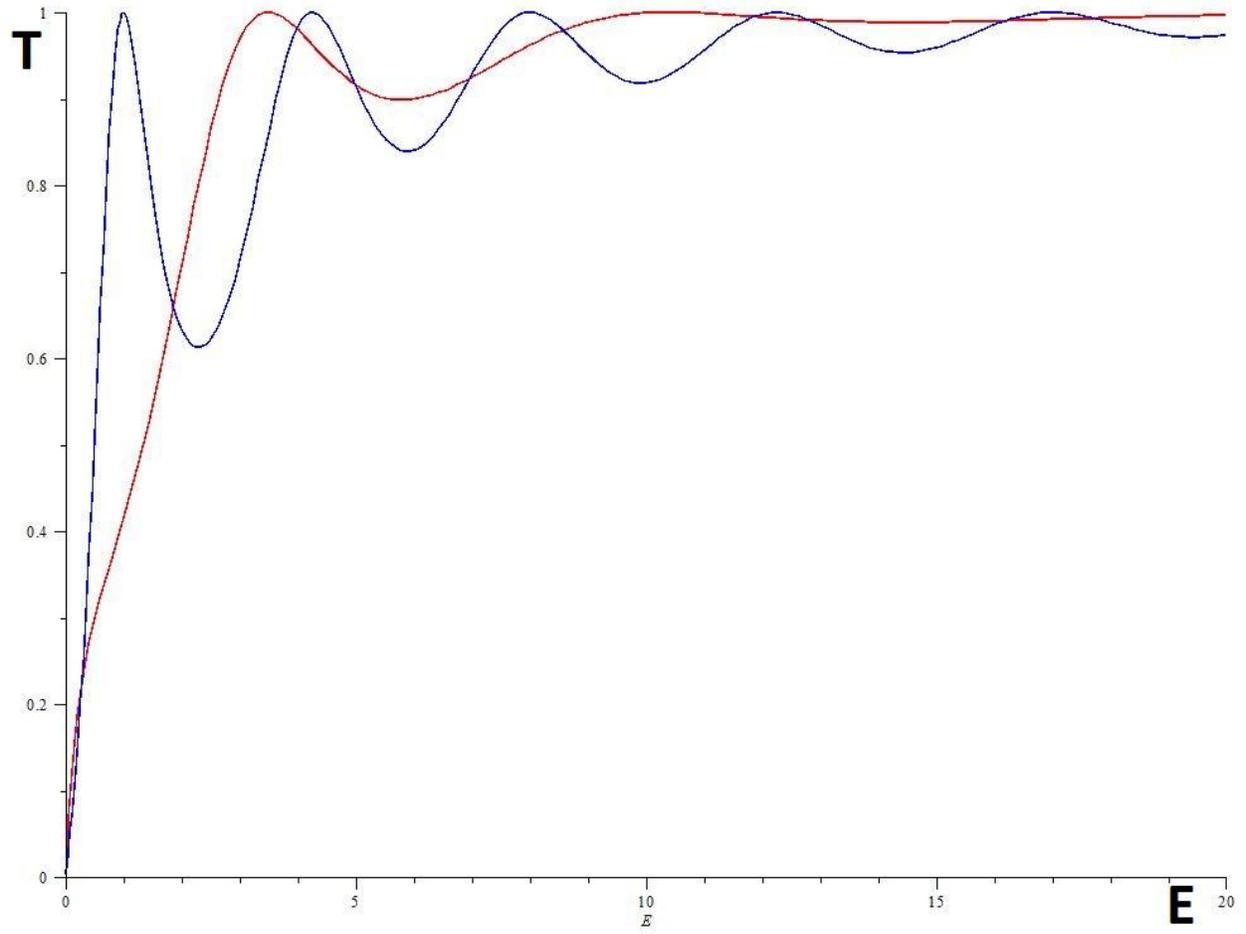

**Figure 5:** Graphs of the transmission coefficient in equation (16) for $\beta = 0$ (blue) and $\beta = 0.01$ (red)